\documentclass[showpacs,preprintnumbers,amsmath,amssymb]{revtex4}

\usepackage{graphicx}
\usepackage{dcolumn}
\usepackage{bm}

\begin{document}

\title{COSMOLOGY OF GRAVITATIONAL VACUUM}

\author{Vladimir Burdyuzha}%
\email{burdyuzh@asc.rssi.ru}
\affiliation{%
Astro-Space Center, Lebedev Physical Institute, Russian Academy of Sciences,
Profsoyuznaya 84/32, Moscow 117810, Russia
}%

\author{Jose Antonio de Freitas Pacheco}
\email{pacheco@obs-nice.fr}
\affiliation{
Observatoire de la Cote d'Azur, bld.de l Observatoire, 06304 Nice Cedex 4, France
}%

\author{Grigoriy Vereshkov}
\email{gveresh@mail.ru}
\affiliation{
Department of Physics, Rostov State University, Stachki 194,Rostov/Don 344104, Russia
}%

\date{\today}

\begin{abstract}
Production of gravitational vacuum defects and their contribution to the energy
density of our Universe are discussed. These topological microstructures (defects) could
be produced in the result of creation of the Universe from "nothing" when a
gravitational vacuum condensate has appeared. They must be isotropically
distributed over the isotropic expanding Universe. After Universe inflation
these microdefects are smoothed, stretched and broken up. A part of them could survive
and now they are perceived as the structures of $\Lambda$-term
and an unclustered dark matter. It is shown that the parametrization noninvariance of the
Wheeler-De Witt equation can be used to describe phenomenologically vacuum
topological defects of different dimensions (worm-holes, micromembranes, microstrings
and monopoles). The mathematical illustration of these processes may be the spontaneous
breaking of the local Lorentz-invariance of the quasi-classical equations of gravity.
Probably the gravitational vacuum condensate has fixed time in our Universe. Besides,
3-dimensional topological defects renormalize $\Lambda$-term.
\end{abstract}

\pacs{}
\maketitle

\section{\label{sec:level1}INTRODUCTION}

\baselineskip = 4mm plus 0.1 mm minus 0.1 mm

Today again we have a new cosmological paradigm. Cosmological acceleration of the
Universe may be explained by a metastable vacuum state in which our Universe is found\cite{B1}.
This metastable vacuum state is probably a state of string landscape. The progress in string
theory\cite{B2},\cite{B3} led to the discovery of a huge quantity of vacuum solutions which were
called string landscape (testing of string theory with CMB can find in the article\cite{B4}).
Research of these ideas combined with ideas of eternal inflation \cite{B5},\cite{B6} have resulted the
development of the idea of inflationary multiverse\cite{B7}. Therefore our Universe may be one
realization of eternaly inflationary multiverse.

The vacuum of our Universe has also a complex structure\cite{B8} and it consists of some subsystems
but here we shall discuss only the gravitation vacuum subsystem. Cosmology of gravitational vacuum
has not been practically discussed elsewhere although the influence of gravity on vacuum has been
considered\cite{B9}. On the other hand cosmology of other vacua is often discussed in the
context of the $\Lambda$-term problem\cite{B10}. Note, that probably dark energy, cosmological constant,
$\Lambda$-term and vacuum energy are the same notion. We are going to consider a gravitational vacuum
condensate which might been produced in the result of creation of the Universe from "nothing"
(this gravitational vacuum condensate is a gravitational vacuum by definition of quantum vacua).

Gravity plays the central role among the fundamental interactions determining
the spacetime structure and it it is an arena of physical reality\cite{B11}.
The question is to find the internal structure of gravitational vacuum starting
from the quantum regime. The quantum regime of gravity has not been satisfactory
explained although some analysis of the problem has been undertaken\cite{B12}, of course.
Besides, there is some analogy between the known vacuum structures and a hypothetical structure
of the gravitational vacuum (it is known that condensates of the quark-gluon type
consist of topological structures - instantons).

The general representation of topological defects can be classified as: 1) 3-dimensional
topological structures ($D = 3$) -worm holes; 2) 2-dimensional topological structures
($D = 2$) -membranes; 3) 1-dimensional topological structures ($D = 1$) -strings; 4)
point defects (singularities) ($D = 0$) - gas of topological monopoles. The full theory of
vacuum defects is absent although we understand that the presence  of defects
breaks the symmetry of a system. The difficult question is to know that type of
symmetry breaking the quantum theory of gravity  gives rise the presence of the
topological defects of the gravitational vacuum. Here it is necessary to include
the experience of vacuum physics.

From a general point of view  we may assume that strong fluctuations of topology
might take place at the Planck scale. Probably there might also exist some stable
structures averaged characteristics of which have been constant or slowly changing
in time. Besides, we must take into account that among possible
parametrization-noninvariant potentials of the Wheeler-DeWitt equation there are ones
topological properties of which may be used for the macroscopic description of
a gas of topological defects (worm-holes, micromembranes, microstrings and monopoles).
This allows us to suggest a hypothesis on which the future theory of quantum gravity
may be solved considering the combination of the problem of quantum topological
structures and the problem of the parametrization-noninvariance of the Wheeler-DeWitt
equation. In the frame of this hypothesis based on mathematical observations and
analogies we suggest the mathematical apparatus of systematic modelling of the
minisuperspace metric and the effective potential permitting us to choose structures
which are interest from the cosmological point of view.

For example, we would like to discuss energy-momentum characteristics of the gravitational
vacuum  which depend on the radius of a closed Universe. Our arguments are only heuristic:

1. This is the simplest mathematical model; \\
2. Our model is agreed with the general property of the Wheeler-DeWitt equation for a closed
isotropic Universe. It is known that the gauge invariance of classical theory of gravity
has an additional aspect - the parametrizational invariance of the regarding choice
of variables on which the gauge conditions are set. In the classical theory a
parametrization and a gauge is a single operation and the division of it on
separate steps is conditional. In the quantum geometrodynamics (QGD) the situation
is different - the basic equations of QGD are gauge invariant but the
parametrization is gauge noninvariant\cite{B13}. Physical consequences of
the parametrizational noninvariance are not general. Some authors\cite{B14} have
suggested to reject this problem and to fix the choice of gauge variables by some
definite way;\\
3. After introduction in cosmology of the quintessence scenario it has been conjectured
that a gravitation vacuum condensate may be considered as a possible factor of
the breaking of the local vacuum Lorentz-invariance.\\
Of course, the gravitational vacuum condensate (as well as its structures) might
appear after the first relativistic phase transition\cite{B15} but in this article
we remain in the frame of Wheeler-DeWitt approach and assume that the phase
transitions are absent.\\
Here it is necessary to stress the following important point related to the notion
of time. As it is known time is one of the coordinates describing a distorted
4-dim Riemann manifold. Appearance of time after production of the gravitational
vacuum condensate fixes the nonconservation of the vacuum symmetry regarding
space-time transformations of the Einstein's theory. Generally speaking the nature
of the full vacuum energy dependence from time is unknown and the mathematical
description of it has a character of discussion.
Models of quintessence\cite{B16} appeal mainly to classical (e.g. scalar) fields.
We shall show that in the quasi-classical approach the quintessence can be mimicked by
mathematical structures arising naturally from the Wheeler-DeWitt's quantum
geometrodynamics. Under the full vacuum energy of the Universe $\Omega_{\Lambda}$=0.7
we understand a "sum" of all vacuum condensates which have been produced after
the relativistic phase transitions during early Universe evolution. These vacuum condensates
might probably compensate the initial vacuum energy $\Omega_{\Lambda}=1$ if the Universe
has been born from "nothing". But this quintessence period of the Universe evolution was not
long. It continues less than one second\cite{B17}.

\section{\label{sec:level1}BASIC STATEMENTS}

The Universe creation is probably a quantum geometrodynamical transition from
"nothing" (the state of "nothing" has geometry of zero volume (a=0)) in
the state of a closed 3-space of small sizes having some particles and fields.
This means that the closed isotropic Universe has been created through tunnelling process\cite{B18}
with a metric:

$$ds^{2} = N^{2} dt^{2} - a^{2}(t) dl^{2} \eqno(1)$$

$$dl^{2} = d\rho^{2} + sin^{2}\rho(d\Theta^{2} + sin^{2}\Theta d\Phi^{2}) \eqno(2)$$

here $N =\sqrt{g_{00}}$ is a gauge variable which is necessary to fix before
the solution of Einstein's equations; and $T^{\mu}_{\nu} = diag(\epsilon, -p,-p,-p)$
is an energy-momentum tensor(TEM). We have restricted our consideration by
only 3-space although a geometrodynamical transition in the state of a closed space of
large dimensions might also be possible. As it is known in the modern epoch the vacuum
energy (which is a dominant component of the Universe) has overcome the curvature and now
the Universe accelerates $\Omega_{\Lambda} \sim 0.7; \Omega_{m} \sim 0.3$ and there is
not contradiction that the Universe has been born close to $\Omega=1$. Besides, the
vacuum energy (if it was positive) was decreasing by jumps due to negative contributions
during the relativistic phase transitions\cite{B15}. On the first stage we would like to
work with the classical theory in which energy-carriers are described by the hydrodynamical
TEM which is locally Lorentz-invariant (generally covariant) and which has
the standard $\Lambda$-term. For a closed Universe Einstein's equations are:

$$R^{0}_{0} - \frac{1}{2}R \equiv 3(\frac{\dot{a}^{2}}{N^{2}a^{2}} +
\frac{1}{a^{2}}) = \ae (\epsilon + \Lambda_{0}) \eqno(3)$$

$$R^{k}_{i} - \frac{\delta^{k}_{i}}{2}R \equiv
\delta^{k}_{i} [\frac{1}{N^{2}}(2 \frac{\ddot{a}}{a} -
2 \frac{\dot{N} \dot{a}}{Na} + \frac{\dot{a}^{2}}{a^{2}}) +
\frac{1}{a^{2}}] = \ae \delta^{k}_{i}(-p + \Lambda_{0}) \eqno(4)$$

$$-R \equiv 6[\frac{1}{N^{2}}(\frac{\ddot{a}}{a} - \frac{\dot{N} \dot {a}}{Na} +
\frac{\dot{a}^{2}}{a^{2}}) + \frac{1}{a^{2}}] = \ae(\epsilon - 3p + 4 \Lambda_{0}) \eqno(5)$$

Note that the equation (5) is the sum of eqs.(3) and (4). Here they have been
written for completeness. In the isotropic Universe with metric (1) and the
mentioned TEM when the Bianchi identity has taken into account the next
equation is obtained:

$$(\epsilon\dot{a}^{4})-\frac{\dot{a}}{a}[(\epsilon-3p)a^{4}]=0        \eqno(6)$$

Assuming that $\epsilon=\epsilon(a)$, p=p(a) we have from eq.(6):

$$p(a)=-\epsilon(a)-\frac{ad\epsilon(a)}{3da}                \eqno(7)$$

If eq.(7) is taken into account then the components of TEM are:

$$T^{0}_{0}=\epsilon(a)$$ $$T^{k}_{i}=\delta^{k}_{i}[\epsilon(a)+\frac{ad\epsilon(a)}{3da}] \eqno(8)$$

These expressions are more controversial in the case of TEM of
a gravitational vacuum. In this case a hydrodynamical velocity can not be used.
Equations (8) may be combined in the noncovariant expression only when:

$$T^{\nu}_{\mu}(GVC)=\delta^{\nu}_{\mu}\epsilon_{GVC}(a)+
\frac{a}{3}(\delta^{\nu}_{\mu}-\delta^{0}_{\mu}\delta^{\nu}_{0})
\frac{d\epsilon_{GVC}(a)}{da} \eqno(9)$$

This expression is not local Lorentz-invariant (here GVC is a gravitational vacuum
condensate). Taking the position of the gravitational vacuum condensate which
is connected with parametrizational noninvariant of the Wheeler-DeWitt equation
the special status of the time index should be understood. In the nonstationary
Universe with noninvariant quantum geometrodynamics with respect to conform
transformations of time the symmetry between space and time is broken. Of course,
the problem is in the parametrizational noninvariance of the Wheeler-DeWitt equation
$H\Psi=0$. Probably existence of this problem indicates that quantum geomerodynamics
of the Wheeler-DeWitt can not pretend to the role of the full theory of quantum gravity.

Now we would like to come back to discussion of equations of the model. Equation (3) is left
without change. In equation (4) the factor $\delta^{k}_{i}$ is cancelled and the relation (7)
is used. Same is done with equation (5). Then we have the next set:

$$3(\frac{\dot{a}^{2}}{N^{2}a^{2}}+\frac{1}{a^{2}})=
\ae[\epsilon_{s}(a)+\Lambda_{0}]  \eqno(10)$$

$$\frac{1}{N^{2}}(2\frac{\ddot{a}}{a}-2\frac{\dot{N}\dot{a}}{Na}+
\frac{\dot{a}^{2}}{a^{2}})+\frac{1}{a^{2}}=
\ae[\epsilon_{s}(a)+\frac{ad\epsilon_{s}(a)}{3da}+\Lambda_{0}] \eqno(11)$$

$$6[\frac{1}{N^{2}}(\frac{\ddot{a}}{a}-\frac{\dot{N}\dot{a}}{Na}+
\frac{\dot{a}^{2}}{a^{2}})+\frac{1}{a^{2}}]=
\ae[4\epsilon_{s}(a)+a\frac{d\epsilon_{s}(a)}{da}+4\Lambda_{0}]  \eqno(12)$$

Here we had introduced the quantum index (s) which numerates matter and vacuum
quantum states. The simple algebraic transformations of equations (11-12) bring to the next equation:

$$\frac{1}{N^{2}}(2\frac{\ddot{a}}{a}-2\frac{\dot{N}\dot{a}}{Na}-
\frac{\dot{a}^{2}}{a^{2}})-\frac{1}{a^{2}}=
\frac{\ae}{3}[\epsilon_{s}(a)+a\frac{d\epsilon_{s}(a)}{da}+\Lambda_{0}] \eqno(13)$$

which is one of motion in an isotropic Universe (equation (10) is one of coupling).
Equation (13) can be brought in the form of total derivative on time of some quantity equal to zero:

$$\frac{d}{dt}(\frac{\dot{a}^{2}}{N^{2}a}+\frac{1}{a}-
\frac{\ae}{3}[a\epsilon_{s}(a)+ a\Lambda_{0}])=0 \eqno(14)$$

Integrating equation (14) and assuming that the constant of integration is zero
it is easy to see that equation (10) is an identity coinciding with the result of integration
of equation (14). But the first stage is not finished yet. The quantum theory of a closed
Universe -quantum geometrodynamics\cite{B19} is based on the Wheeler idea of
a superspace. This idea includes the manifold of all possible geometries of 3-space,
matter and field configurations where the Universe wave function is defined.
Before quantization of the classical theory it is necessary to impart the form of
Lagrange and Hamilton theory with couplings. We are able to quantize only a Hamilton
theory but the construction of a Hamilton theory precedes the construction of a Lagrange
one.

Evidently that two variables must be presented in Lagrange formulation of the theory:
the dynamical variable $a(t)$ relating to an equation of motion and some Lagrange
multiplier $ \lambda = \lambda(t)$ relating to a coupling equation. From an
infinite number of possibilities of inserting of a Lagrange multiplier we
have chosen a particular variant related with quantum geometrodynamics. In quantum
geometrodynamics a problem arises which has not the classical analogy: it is necessary
to formulate the procedure of operators ordering on a generalized momentum and
a coordinate. It is assumed that the procedure of ordering must be based on the
covariance principle of the Wheeler-DeWitt equation in the Wheeler superspace whose metric
is $ \gamma(a)$.

The explicit form of function $ \gamma (a)$ is not fixed  but this conception
allows us to understand in which terms the problem of the parametrizational
invariance is formulated. The above mentioned program can be carried out only when:
$N = \frac{\lambda a}{\gamma(a)}$ where $ \lambda$ is the Lagrange multiplier,
$ \gamma(a)$ is the metric of a superspace. Then we have:

$$3\left(\frac{\gamma^{2}{\dot a}^{2}}{\lambda^{2} a^{4}} + \frac{1}{a^{2}}
\right) = \ae \left[ \epsilon_{s}(a) + \Lambda_{0} \right] \eqno(15)$$
\begin{widetext}
$$
\frac{\gamma^{2}}{\lambda^{2} a^{2}} \left( 2 \frac{\ddot {a}}{a} -
2 \frac{\dot {\lambda} \dot {a}}{\lambda a} + 2 \frac{\dot {\gamma} \dot {a}}
{\gamma a} - 3 \frac{\dot {a}^{2}}{a^{2}} \right) -
\frac{1}{a^{2}} = \frac{\ae}{3} \left[ \epsilon_{s}(a)+a \cdot
\frac{d \epsilon_{s}(a)}{da}+ \Lambda_{0} \right] \eqno(16)$$
\end{widetext}
The mathematically equivalent equations  can be obtained from the variational
principle determined by some effective action. The gravitational
part of this action is the known expression for an isotropic Universe:

$$S_g=\int \left(\frac{1}{2\ae} R+ \Lambda_{0} \right){\sqrt {-g}}d^4x \eqno(17)$$

Some standard operations have to be performed with this expression: transformation
of this expression to the quadratic form with respect to the generalized velocity $\dot a$
by excluding the total derivative; integration over the volume of the closed Universe $V=2\pi^2a^3$;
insertion of time parametrization. The effective action of matter and radiation
is added to the received result using Rubakov-Lapchinsky prescription\cite{B20}. This
prescription is enough simple: the energy density of matter $\epsilon_{s}(a)$ depending
on radius of the Universe as well as the $ \Lambda$-term have the mathematical status of
an effective potential energy. Therefore, in order to obtain the right expression it
is necessary to replace $\Lambda_{0}$ on $\Lambda_{0} + \epsilon_{s}(a)$.
The final expression for the effective action is:

$$S_{\gamma} \{a, \lambda\}= \int L_{\gamma} (a, \lambda) dt,$$ \qquad
$$L_{\gamma} (a, \lambda) = \frac{6 \pi^{2}}{\ae} \frac{1}{\lambda}
\gamma(a) \dot{a}^{2} - \lambda \frac{aU(a)}{\gamma(a)} \eqno(18)$$

where

$$U(a) = \frac{6 \pi^{2}}{\ae} a - 2 \pi^{2} a^{3} [\epsilon_{s} (a) +\Lambda_{0}]$$

is the total effective potential energy  accounting for topology of the closed
Universe, the standard $\Lambda$-term and the matter. The variation of action with respect to
$\lambda (t)$ gives:
\begin{widetext}
$$
\frac{\delta S_{\gamma} \{a, \lambda\}}{\delta \lambda}=
\frac{\partial L_{\gamma} (a, \lambda)}{\partial \lambda}=
-\frac{2 \pi^{2}}{\ae \gamma}\left(\frac{3\gamma^2{\dot a}^2 }{\lambda^2}+3a^2-\ae a^4[\epsilon_{s}(a)+
\Lambda_{0} \right) = 0 \eqno(19)
$$
\end{widetext}
which is equivalent to the coupling equation. The variation with respect to the
dynamical variable $a(t)$ gives the Lagrange equation:

$$\frac{\delta S_{\gamma} \{a, \lambda\}}{\delta a}=-\frac{d}{dt}\frac {\partial
L_{\gamma} (a, \lambda)}{\partial \dot a}+\frac {\partial L_{\gamma} (a,\lambda)}{\partial a}=0 \eqno(20)$$

After some transformations of this equation we have:
\begin{widetext}
$$
\frac{d}{dt} \frac{\partial L_{\gamma}(a, \lambda)}{\partial \dot{a}}
- \frac {\partial L{\gamma} (a, \lambda)}{\partial a} \equiv
\frac{6 \pi^{2} \lambda}{\ae \gamma} \left\{ \frac{\gamma^{2}}{\lambda^{2}}
\left( 2 \ddot {a} +2 \frac{\dot {\gamma} \dot{a}}
{\gamma} - 2 \frac{\dot{\lambda} \dot{a}}{\lambda} - 3 \frac{\dot{a}^{2}}{a} \right)
- a - \frac{\ae a^{3}}{3} (\epsilon_{s}(a) +a \frac{d \epsilon_{s}(a)}{da} +
\Lambda_{0}) + J \right\} = 0 \eqno(21)
$$
\end{widetext}
where

$$J=\frac{2 \pi^{2} \lambda}{\ae \gamma}\cdot \frac{d\ln (a^{-3} \gamma
(a))}{da} \left( \frac{3 \gamma^{2} \dot {a}^{2}}{\lambda^{2}} + 3a^2-\ae
a^4[\epsilon_{s}(a)+\Lambda_{0}]\right)$$.

These equations give the set of equations of the Lagrange model. Of course, these
equations must be considered in a combination. It is not difficult to see that for $J=0$
the last equation is mathematically equivalent to the combination of the Einstein's equations
discussed before. Thus we have proved that this model gives us
the Lagrange method of the description of an isotropic Universe, energy-carriers
of which are described by the functions of the scale factor $\epsilon_s(a),\;p_s(a)$.
In the classical theory this result has only methodical character.
However, we want to transfer these ideas to the quantum geometrodynamics where
the Hamilton formulation is necessary. The Hamilton model can be built on the basis
of a Lagrange model. Note that introduction of the function $\gamma(a)$ is the operation
of parametrization. The index $\gamma$ shows that the action and the Lagrangian correspond to
a definite parametrization. The Hamiltonian of our set is built by the standard procedure:

$$H\Phi_{s}=0 \eqno(22)$$

$$H = P \dot{a} - L = \lambda (\frac{\ae}{24 \pi^{2}} \frac{1}{\gamma} P^{2}+
aU(a)) \eqno(23)$$

where $P = \frac{\partial L}{\partial \dot{a}} = \frac{12 \pi^{2}}{\ae}
\frac{\gamma}{a} \dot{a}$ is a generalized momentum. Note that the Wheeler-DeWitt
theory also has the parametrization problems. The commutation problem
implies that the operator $p$ is not uniquely defined:
$\hat{p} =-i \hbar \frac{\partial}{\partial a} +f(a)$ where f(a) is an arbitrary
function. The second problem is the ordering of operators in the Hamiltonian (this
is significant for any nontrivial function $\gamma(a)$). The partial solution of these problems is
proposed in the frame of hypothesis of covariant differentiation in a curved
space. According to this hypothesis the Wheeler-DeWitt equation has the form:
\begin{widetext}
$$-\frac{\ae \hbar^{2}}{24 \pi^{2}} \frac{1}{\sqrt{\gamma}} \frac{d}{da}
\frac{1}{\sqrt{\gamma}} \frac{d \Phi_{s}(a)}{da} + \frac{1}{\gamma}
[ \frac{6 \pi^{2}}{\ae} a^{2} - 2 \pi^{2} a^{4} (\epsilon_{s}(a) + \Lambda_{0})]
\Phi_{s} (a) = 0 \eqno(24)$$
\end{widetext}
and the wave function of the Universe satisfies the condition:

$$\int^{\infty}_{0} \sqrt{\gamma} \; da \Phi^{\star}_{s}(a) \Phi^{\star}_{s'}
(a)= \delta (s - s^{'}) \eqno(25)$$

where  $\delta(s - s^{'})$ is the delta function depending on the particular
properties of the solutions of the Wheeler-DeWitt equation. Unfortunately the hypothesis of covariant
differentiation does not solve completely the parametrization problems of the theory.
To stress the multiplicative redefinition of the wave function: $\Phi_{s} (a) = \gamma^{1/4} \Psi_{s} (a)$
is used and then the Wheeler-DeWitt equation is rewritten in the form:

$$(\frac{\ae \hbar}{12 \pi^{2}})^{2} \frac{d^{2} \psi_{s}}{d a^{2}} +
[a^{2} - \frac{a^{4}}{3} (\ae \epsilon_{s} (a) + \ae \Lambda_{0} +
\ae \epsilon_{GVC} (a))] \Psi_{s} (a) = 0 \eqno(26)$$

where

$$\epsilon_{GVC}(a) = \frac{\ae \hbar^{2}}{192 \pi^{2}} \frac{1}{a^{4}}
(\mu^{''} - \frac{1}{4} (\mu^{'})^{2}) \eqno(27)$$

Here the sign ${'}$ is a derivative of the parametric function $\mu(a) =
ln\gamma(a)$ with respect to the scale factor. All parametrizational noninvariant effects are
collected in the function $\epsilon_{GVC} (a)$ which we call
the density of energy of a gravitational vacuum (a gravitational vacuum condensate
$(GVC)$). As it is easy to see that parametrizational noninvariant effects are
clearly the quantum ones, $\epsilon_{GVC} \sim \hbar^{2}$. The parametrization noninvariant
contributions have not a obvious physical meaning since we do not know well their  physical
nature. These contributions have been arisen by reason of nonconservation
of the classical symmetry on the quantum level. The experience of the modern quantum field
theory tells that when a symmetry is broken a vacuum state is also changed. Therefore
we called the parametrizational noninvariant contributions as the
density of a GVC energy.

However, general symmetric arguments  do not have the clear physical connection to
the vacuum energy. In this situation the examples of QCD are useful. In the quantum
theory classical conformal  and chiral symmetries do not conserve resulting in
appearance of a quark-gluon condensate. Probably particular vacuum topological
structures exist in a gravitation vacuum and they are the consequence of the
parametrizational noninvariance of the quantum geometrodynamics.

\section{\label{sec:level1}COSMOLOGICAL APPLICATION}

In cosmology this means that properties of the topological microscopic defects
of space on average are isotropic and homogeneous (isotropisation in the brane
gas cosmology is also a natural consequence of the dynamics\cite{B21}).
They are contained in the function $\mu(a)$. We propose that all topological quantum
defects with $D \ge 1$ have the typical Planck size. On this reason breaking up
of the defects resulting a change of their number in a variable volume $V = a^{3} (t)$
should take place. The number of defects in this volume can be estimated as:

$N_{D} \sim (\frac{a}{l_{pl}})^{D}, \;\;\; l_{pl} = (G \hbar)^{1/2} =
\frac{(\ae \hbar)^{1/2}}{\sqrt{8 \pi}},$  here $c = 1$.

According to our estimates we expect that the energy density of
the system of topological defects contains a constant part corresponding to
worm-holes and members of the types  $1/a^{3}; 1/a^{2}; 1/a$  corresponding
to a gas of point defects, micromembranes and microstrings. Besides, the function
$\epsilon_ {GVC}(a)$ must contain additional members describing interactions of
microdefects between each other. This corresponds to the
next choice of the function $\mu(a)$:

$$\mu(a) = c_{0} ln a + c_{1} a + \frac{1}{2} c_{2} a^{2} + \frac{1}{3}
c_{3} a^{3}, \;\;\; c_{i} = const \eqno(28)$$

Then we have:
\begin{widetext}
$$\Lambda_{0} + \epsilon_{GVC} (a) = \Lambda_{0} - \frac{\ae \hbar^{2}}
{768 \pi^{2}}\; c^{2}_{3} + \frac{\ae \hbar^{2}}{192 \pi^{2}}
[- \frac{1}{2} \; c_{2} c_{3} \frac{1}{a} - (\frac{1}{4} c^{2}_{2} +
\frac{1}{2} c_{1} c_{3}) \frac{1}{a^{2}} +$$
$$+ (2 c_{3} - \frac{1}{2} c_{1} c_{2} - \frac{1}{2} c_{0} c_{3}) \frac{1}{a^{3}} +
(c_{2} - \frac{1}{4} c^{2}_{1} - \frac{1}{2} c_{0} c_{2}) \frac{1}{a^{4}} -
\frac{1}{2} c_{0} c_{1} \frac{1}{a^{5}} - (c_{0} + \frac{c^{2}_{0}}{4})
\frac{1}{a^{6}}] \eqno(29)$$
\end{widetext}
Here as indicated above the last three members can be interpreted as the energy of gravitational
interaction of defects between each other but their discussion is not the case
since the quasi-classical dynamics is correct in the region of the large scale factor a.
Note also that 3-dimensional topological defects renormalize $\Lambda$-term.
Observable value of $\Lambda$-term should be:

$$\Lambda = \Lambda_{0} - \frac{\ae \hbar^{2}}{768 \pi^{2}} c^{2}_{3} \eqno(30)$$

while the term $\frac{1}{a^{3}}$ may be relevant to dark matter(DM). This gives
the limitation on parameters of the function  $\mu(a)$:

$$\frac{1}{3} \; l^{4}_{pl}\; [2 c_{3} (1 - \frac{c_{0}}{a}) - \frac{1}{2} c_{1}
c_{2}] = \ae M \eqno(31)$$

where $M$ is a mass in volume $a^{3}$. As it is known the Wheeler-DeWitt quantum
geometrodynamics is an extrapolation of quantum-theoretical concepts on the scale of the
Universe as whole. The initial state of the Universe from QGD point of view was located
in the region of small values of the scale factors in a minisuperspace. From the
classical point of view the initial state of the Universe
is a structureless singular state. Here we must postulate the defect creation
of the Universe if it has been born from "nothing".  After the creation of defects
probably in our Universe the stage of quick expansion (inflation) has taken
place. In the result defects have been smoothed, stretched and broken up. Some defects
could have been left and give rise to the $\Lambda$-term and the unclustered DM. Physics
of topological defects arising during the phase transitionshas has in detail been discussed
by T.Kibble\cite{B22}.

\section{\label{sec:level1}CONCLUSION}

We have discussed a possible quantum modification of the minisuperspace Wheeler-DeWitt
equation due to operator ordering ambiguities and have proposed to interpret it as
being due to various topological defects. We have taken into account that among
parametrizational noninvariant potentials of this equation there are ones which have
properties suitable to a macroscopic description of a gas of topological
defects (worm-holes, micromembranes, microstrings and monopoles). Parametrizational
noninvariant members arisen from the Wheeler-deWitt theory have not analogies in the local
quantum theory of field. Their origin are connected with a global quantization. An
appropriate degree of freedom (a scale factor of the closed Universe) does not satisfy
to the asymptotical conditions used in the formalism of the local quantum theory of
field. And what is more, in the local quantum theory of field this degree of freedom
as a physical quantum variable is completely absent.  Probably in the future
theory of quantum gravity the problem of topological structures of gravitational vacuum
and parametrizational noninvariance of the Wheeler-DeWitt equation must be solved together
(authors\cite{B23} even attempted to parametrize the state equation of the dark energy).
Also we have shown that the quasi-classical corrections ( proportional to
$\sim \hbar^{2}$) are completely defined by a superspace metric (that is $\gamma(a)$).
This means that these corrections in the theory of gravity are not entirely
determined by physics of the 4-dim space-time. In the frame of quantum geometrodynamics
a part of the corrections having an influence on the evolution of the Universe in 4-dim
is determined by physics of a superspace (that means that they come from another level
of a theory). If we interpret these corrections as defects then this means that defects
appear as a result of interaction of universes in this superspace. The development of
these ideas may be realized in the frame of tertiary quantization.

Besides, we note that the property of the Lorentz invariance attributed to 4-manifold
connects 1-dim time and 3-dim space. Here 3-dim defects (worm-holes) give a contribution
in the Lorentz-invariant $\Lambda$-term. Quantum topological defects with D=0,1,2
give the Lorentz-noninvariant contributions in vacuum TEM. Thus, we emphasize that
topological defects of gravitational vacuum are quantum structures produced at
the Planck epoch of the Universe evolution (the Lorentz invariance at the Planck scale
must probably be modified \cite{B24}). Besides, the important moment is also understanding
that the gravitational vacuum condensate has fixed time in our Universe. Topological
defects of the gravitational vacuum give rise to the  $\Lambda$-term
and the unclustered dark matter. Probably these ideas allows us to improve our
understanding of main components of the Universe.

Finally we conclude that in the frame of the proposed hypothesis (the defect
microstructure of space as a carrier of quintessence energy) a set of serious
problems arises with no obvious solutions. At first it is the problem of evolution
of the defects on different stages of the cosmological expansion. It is natural
to suggest that in the frame of any scenario of the Universe creation from "nothing"
the initial density of microdefects had the Planck value. The quick expansion of
the Universe embryo with characteristic time of the order of the Planck time
has led to destruction of a majority of microstructures and to
transition of their energy density to the energy density of elementary particles.
It is not excluded that similar processes can lead to some characteristic peculiarities
in the spectrum of relict perturbations of density.
Besides, equations which have been used in this research pretend only on a
phenomenological description of the energy density of an extremely small amount of
the microdefects which have been left after a violent period of their relaxation during
the Planck epoch of the Universe evolution. In the post-Planck epoches when
the characteristic time of expansion is much large than the Planck time scale by
many orders of magnitudes one may hope that a more slow expansion did not lead to
additional destruction of the left small amount of microdefects.

The preliminary version of this article can be found in\cite{B25}. Other time we underline
that quintessence period of the Universe evolution might continue less than one second after
it creation. In this period of the Universe evolution a vacuum component was changing by jumps
during phase transitions. Vacuum condensates of quantum fields carried a negative contribution
in its initial positive density energy\cite{B17}.

The more general cosmological researches of vacuum and a new cosmological paradigm based on vacuum
string landscape and eternal inflation in multiverse can be found in the articles\cite{B26}.


\begin{thebibliography}{26}
\bibitem{B1} S.Kachru, R.Kallosh, A.Linde and Triverdi Phys.Rev D68, 046005 (2003).
\bibitem{B2} A.Moloney, E.Silverstein, A.Strominger hep-th/0205316.
\bibitem{B3} L.Susskind hep-th/0302219.
\bibitem{B4} R.Kallosh and A.Linde hep-th/0704.0647, 5 Apr 2007.
\bibitem{B5} A.Vilenkin Phys.Rev.D27, 2848, (1983).
\bibitem{B6} A.Linde, D.Linde, A.Mezhlumian Phys.Rev D.49, 1783 (1994).
\bibitem{B7} J.Garriga, D.Schwartz-Perlov, A.Vilenkin and S.Winitzki hep-th/0509184, 27 Nov 2005.
\bibitem{B8} V.Burdyuzha and G.Vereshkov Astrophys.Space Sci.305, 235 (2006).
\bibitem{B9} S.Coleman and F.De Luccia, Phys.Rev.21, 3305 (1980).
\bibitem{B10} M.Bordag, Phys.Rev.D67, 065001 (2003); P.Jaikumar and A.Mazumbar,
Phys.Rew.Lett 90, 191301 (2003); A.Gomberoff, M.Henneaux, C.Teitelboim hep-th/0501152,v.1, 19 Jan 2005.
\bibitem{B11} S.V.Babak and L.P.Grishchuk, gr-qc/0209006; A.Zee hep-th/0309032.
\bibitem{B12} G.Preparata, S.Rovelli, and S.-S.Xue, gr-qc/9806044; S.Carlip, Rep.Prog.Phys.64, 885 (2001);
T.Banks, hep-th/0305206 v.1; C.Kiefer, gr-qc/0508120, 21 Sep. 2005; D.Giolini and C.Kiefer gr-qc/0611141 27 Nov.2006.
\bibitem{B13} M.J.Duncan, Preprint UMH-TII-916/96; V.A.Savchenko, T.A. Shestakova, G.M.Vereshkov Intern.J.Mod.Phys. 15, 3207 (2000);
Gravity and Cosmology 25, 18 (2001).
\bibitem{B14} S.W.Hawking and D.N.Page, Nucl.Phys.B264, 185 (1986).
\bibitem{B15} V.Burdyuzha, O.Lalakulich, Yu.Ponomarev, and G.Vereshkov, Preprint of Yukawa Institute
for Theoretical Physics (YITP-98-50) and gr-qc/9907101.
\bibitem{B16} R.R.Caldwell, R.Dave, P.J.Steinhardt, Phys.Rev.Lett. 80, 1582 (1998); R.R.Caldwell and E.V.Linder, Phys.Rev.Lett 95, 141301 (2005).
\bibitem{B17} V.Burdyuzha, Phys.Lett., submitted (2007).
\bibitem{B18} V.Burdyuzha, O.Lalakulich, Yu.Ponomarev, and G.Vereshkov, Phys.Rev.D55, 7340R, (1997).
\bibitem{B19} B.DeWitt, Phys.Rev.160, 1113 (1967).
\bibitem{B20} V.G.Lapchinsky and V.A.Rubakov, Acta Phys.Polonica B10, 1041 (1979).
\bibitem{B21} S.Watson and R.Branderberger, Phys.Rev.D67, 043510 (2003).
\bibitem{B22} T.Kibble "Topological Defects and Non-Equilibrium Dynamics of Symmetry Breaking Phase Transitions"
Eds.Yu.Bunkov.H.Godfrin, Kluwer Academic Publishers, The Netherlands (2000).
\bibitem{B23} P.S.Corasaniti and E.J.Copeland, Phys.Rev.D67, 063521 (2003).
\bibitem{B24} J.Magueijo and L.Smolin, Phys.Rev.D67, 044017 (2003); T.Jacobson, S.Liberati
and D.Mattingly, Nature 424, 1019 (2003).
\bibitem{B25} V.Burdyuzha, J.A. de Freitas Pacheco, G.Vereshkov , gr-qc/0312072.
\bibitem{B26} J.Garriga, A.Guth, A.Vilenkin hep-th/0612242,v.1,21 Dec (2006); S.H.Henry Tye hep-th/0610221 (14 Nov.2006);
R.Bousso, B.Freivogel, I-Sheng Yang hep-th/0606114v.2,15 Sep.(2006); L.Mersini-Houghton hep-th/0512304, Apr.25 (2007);
A.Linde hep-th/0611043,v.3,28 Jan (2007).

\end{thebibliography}
\end{document}